# Ion Trap Geometry

Evgeny V Krylov[1]

## Abstract

This work aims to find ion trap geometry for a high-quality ion trap mass analyzer that minimizes mass shift and is easy to fabricate. The theoretical procedure of the ion trap mass shift evaluation is developed, which provides a quantitative criterion of the mass shift (ion ejection delay in comparison with ideal ion trap) for the correct comparison of the different ion trap geometries. This procedure is validated by comparison with known experimental facts about existing mass analyzers: 3D hyperbolic ion trap (Finnigan), 2D hyperbolic ion trap (Thermo), Rectilinear Ion Trap (Purdue), and quadrupole mass filter. The developed theoretical procedure optimizes known ion trap geometries for low mass shift. Applying this procedure to various ion trap geometries reveals that in terms of the mass shift existing ion traps are close to the optimum in a framework of chosen geometry. Alternative ion trap geometry of very low mass shift, simple in fabrication, and satisfying the requirement of high mechanical accuracy is proposed. The proposed alternative ion trap geometry provides a lower ejection delay than linear ion traps with hyperbolic rods.

## Introduction

Ion trap mass spectrometry is increasingly important in modern instrumental analysis. Capabilities for identifying and quantifying both large and small compounds facilitate the investigation of complex chemical or biochemical systems with high sensitivity and versatility. The attractiveness of ion trap mass spectrometry can also be attributed to the high-quality analytical performance achieved using a relatively simple mass analyzer device. In particular, the ability to perform tandem mass spectrometry using a single analyzer instrument is a major advantage.

Electrodynamics ion traps date back to the pioneering work of Wolfgang Paul[1] who first described the three-dimensional electric quadrupole field established by three electrodes with hyperbolic surfaces and their ion-trapping capabilities. The ring electrode is supplied with a megahertz radio (RF) frequency voltage, and the two endcap electrodes are normally grounded. Different ion trap geometries have evolved along with half a century's development, as modifications on the original Paul design either for performance improvement or simplifications for specific applications.

### Ion Trap theory

Ion traps are dynamic instruments in which time-dependent forces influence ion trajectories. Ions in the electric fields that arise from applying a quadrupolar potential experience strong focusing (the restoring force increases linearly with displacement from the center of the device and drives the ions back toward the origin). The motion of ions in such fields is described mathematically by the solutions to a second-order linear differential equation (Mathieu's equation).

The equations for the electrode surfaces in the hyperbolic ion trap are $r^2 - 2z^2 = r_0^2$ for the ring electrode and $r^2 - 2z^2 = z_0^2$ for the endcap electrodes. Potential inside a hyperbolic ion trap is given by $\Phi(r,z,t) = \frac{U+VCos(\omega t)}{r_0^2+2z_0^2}(r^2 - 2z^2 + 2z_0^2)$, where $U + VCos(\omega t)$ is the voltage applied to the ring electrode. The components of the ion motion in the radial (r) and the axial (z) directions may be considered independently, so for the axial direction we can write the ion movement equation (similar to Mathieu's equation) $\frac{d^2z}{dt^2} + \left[\frac{4eU}{m(r_0^2+2z_0^2)} + \frac{4eVCos(\omega t)}{m(r_0^2+2z_0^2)}\right]z = 0$, which can be solved in terms of the stability diagram in axes $a_z = -\frac{16eU}{m(r_0^2+2z_0^2)\omega^2}$ and $q_z = \frac{8eV}{m(r_0^2+2z_0^2)\omega^2}$ for z-direction; for r-direction $a_r$=-$a_z$/2 and $q_r$=-$q_z$/2. Trapping parameter $\beta_u = \sqrt{a_u + \frac{q_u^2}{2}} < 1$ (simplified approximation) defines stability regions in (a;q) coordinates.

---

[1] E-mail address: great418@gmail.com

The electric field time-averaged effect is essentially a pseudopotential with a linear restoring force pushing ions toward the trap center, hence the ion's motion could be approximated as oscillating in a parabolic pseudopotential well. Well depth in the z and r dimensions is given by $D_z = Vq_z/8$ ; $D_r = Vq_r/8$. The maximum density of singly-charged particles that can occupy an ideal Hyperbolic Ion Trap (HIT) is estimated as[2] $N_{max} = 3V/4\pi m\omega^2 r_0^4$. HIT at modest trapping voltages stores $10^2$–$10^5$ ions.

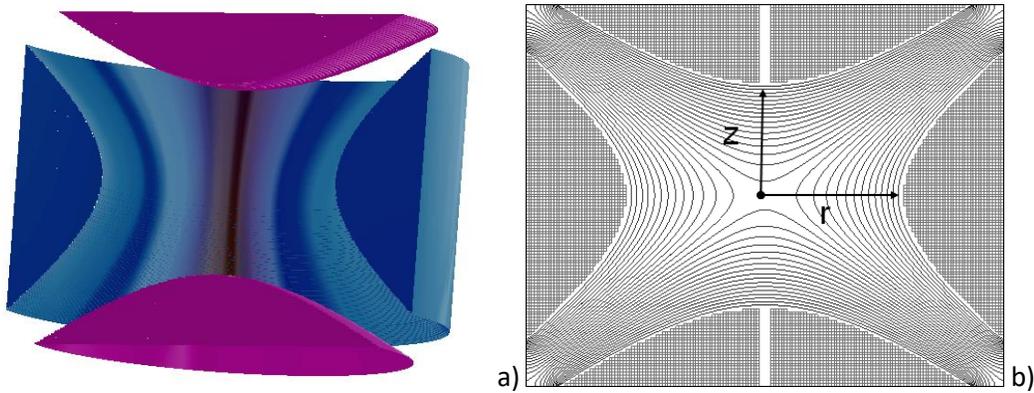

**Figure 1. Ideal Hyperbolic Ion Trap (HIT) a) design; b) cross-section, field lines, and axes.**

Three main disadvantages of the Hyperbolic Ion Trap led to exploring alternative geometries. These disadvantages are the low ion capacity of the three-dimensional (3D) HIT, the difficulty in implementing hyperbolic electrodes of high accuracy, and the mass shift associated with non-linear resonance ejection. We address these problems below.

*Ion capacity*

The analytical performance of ion traps increases with the number of trapped ions provided that space charge effects are minimized. 3D Hyperbolic Ion Traps have inherently limited ion trapping capacity, due to the three-dimensional nature of the ion trapping which confines trapped ions to a small volume at the center of the device. It restricts the actual number of trapped ions to only a few hundred. Moreover, ions injected through the endcap electrode holes experience a strong RF potential, which restricts the trapping efficiency for externally injected ions. A two-dimensional (2D) Linear Ion Trap (LIT) was developed for larger trapping capacity and higher trapping efficiency[3].

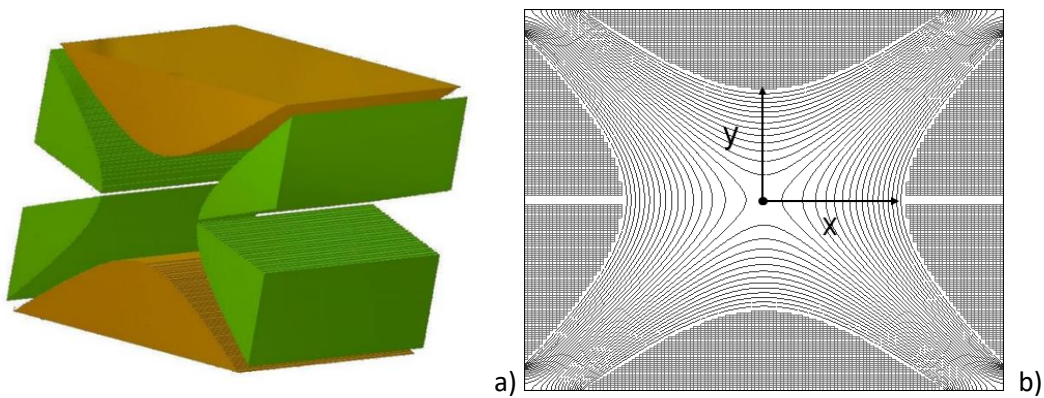

**Figure 2. Linear hyperbolic ion trap (LIT): a) design; b) cross-section, field lines, and axes.**

Its structure is derived from the quadrupole mass filter with two additional endcap electrodes. The RF voltage applied to the pairs of hyperbolic rods and the DC voltage applied on endcap electrodes, establish a cylindrical trapping volume considerably larger than a spherical volume generated by a 3D Hyperbolic Ion Trap of similar size. It results in a significantly increased trapping capacity (10…100 times) fundamentally associated with trapping along a line vs. at a point. Ions are injected into a linear trap along the axial direction and thus not subjected to a direct RF field, which leads to enhanced trapping efficiency for external ion injection.

### High geometric and positioning accuracy

Mass accuracy, resolution, and low ion losses require low field distortions. High mechanical accuracy at the micron level is essential to minimize field imperfection.

Let's consider a linear ion trap with non-parallel electrodes to estimate how mechanical accuracy affects the mass spectrometer resolution. A small angle α between slotted electrodes characterizes non-parallelism. If the electrodes are non-parallel, ejection conditions depend on the ion position along z-axe, so the ions of slightly different mass are ejected simultaneously. That results in the mass spectrum peak broadening. Ejected mass ($m_{ej}$) depends on the distance between electrodes (x) as $m_{ej} \sim x^{-2} \approx x_0^{-2}(1 - 2\,\alpha z/x_0)$, where $x_0$ is the average distance between electrodes; $\alpha z$ is a small deviation of the distance between electrodes along the z-axe. It results in the MS peak broadening by $\Delta m = 2m\,\alpha \Delta z/x_0$. So, the maximal reachable MS resolution is $Q_{max} \approx x_0/\alpha z_0$, where $z_0$ is the electrode length, substituting typical values (Q=1000, $x_0$=1cm) yields accuracy requirements of 10 μm.

From an approximate solution to the equations of ion motion in distorted fields, Titov[4] concluded that a mechanical error of 10 μm increases ion losses by a factor of nearly 10. Dawson[5] stated that quadrupole rods are said to be manufactured to tolerances in radius of $10^{-5}$ cm and parallelism of $10^{-4}$ cm. The electrodes' positioning, torque, and surface quality are not so critical.

The difficulties in implementing hyperbolic electrode structures with high accuracy led to exploring a simpler alternative, the Rectilinear Ion Trap[6] (RIT). This mass analyzer consists of two pairs (x and y) of planar electrodes mounted in parallel. The RF voltage is applied between the x and y electrode pairs to form a quasi-quadrupolar trapping field in the x-y plane.

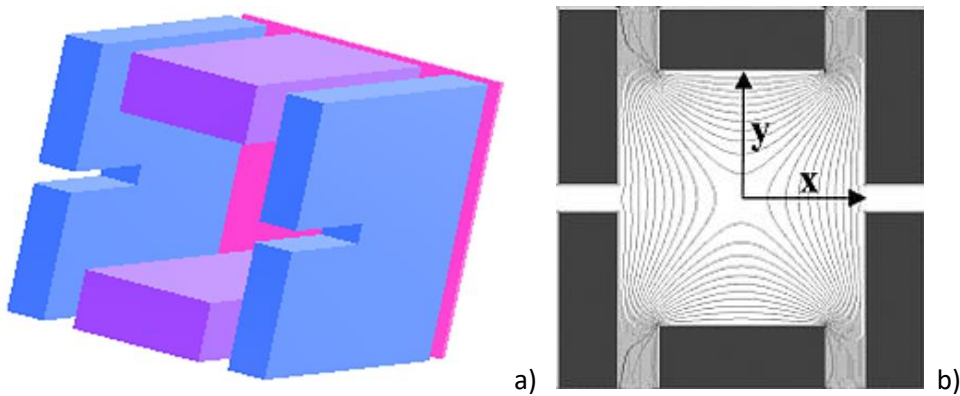

a) b)

**Figure 3. Rectilinear ion trap (RIT): a) geometry; b) cross-section, field lines, and axes.**

### Mass shift

One of the most important features of any mass spectrometer is that it consistently and reproducibly indicates the correct value of the mass/charge ratios of the ions formed from the sample being investigated. If the mass value is "shifted" from its true value for some reason the MS instrument is useless. Mass measurement error occurring for ions of the same mass (for various reasons) is called mass shifts and has been recognized as the major factor influencing MS performance. The mass shift can be as large as 0.1%.

During the commercialization of the 3D Hyperbolic Ion Trap by Finnigan Corp in 1984, staff members realized that some compounds did exhibit either positive or negative mass shifts. To fix the problem Paul Kelley decided to displace outward both end-cap electrodes by the extraordinary amount of 10%. All Finnigan's ion trap mass spectrometers (ITD 700, ITD 800) used such ion traps. Varian and Thermo Scientific faced the same mass shift problem with their 2D Linear Ion Traps MS instruments (Saturn GC/MS and LTQ). They had solved it similarly: moving the ejection electrodes outward.

In 2003, Plass[7] found that mass shift results from an ejection delay caused by the field imperfections near the holes in the end-cap electrodes within ion traps of non-optimized geometry. Below is a brief overview of the ion trap mass shift theory.

## Resonant ion ejection

Ion trap mass spectrometers employ a resonance technique (boundary or resonance ejection scans) to eject ions from the ion trap. The ideal ejection process would allow ions to be ejected unidirectional through the ejection holes as quickly as possible, so the mass spectrum is acquired rapidly, with stable well-resolved MS peaks. Delayed ejection can result in MS peak broadening and/or mass shift within the mass spectrum. By decreasing the scan rate of the ion trap it is possible to improve the peak width, but this method has negative side effects: high ion losses (hence low sensitivity) and longer scan time.

In an ideal quadrupolar field the magnitude of the field strength increases linearly with distance from the trap center, so the ion ejection is a linear resonance process. It means that ion oscillation eigenfrequency ($\omega_0$) is independent of the oscillation amplitude ($X$); the ion motion in the x and y directions is uncoupled; ion oscillation eigenfrequency doesn't change during the ejection, so the ions are ejected as quickly as possible.

However, a purely quadrupolar field inside a real ion trap can't be implemented because of the field distortion due to the ejection holes; and imperfection of the electrodes (geometry, position, and surface). So, the magnitude of the field strength increases non-linearly with distance from the trap center, which results in non-linear resonance. It means that ion oscillation eigenfrequency depends on the oscillation amplitude, $\omega_0 = f(X)$. During the ejection process, the ion oscillation amplitude increases, and hence, the ion eigenfrequency deviates from the exciting electric field frequency, disturbing the resonance conditions. Non-linear resonance gives rise to a prolonged delay in ion ejection in the mass-selective instability scan and coupling between axial and radial motion (this effect is small during typical ion trap operation). In mass spectrometry, prolonged ejection delay results in mass shift, peak broadening, and ion losses.

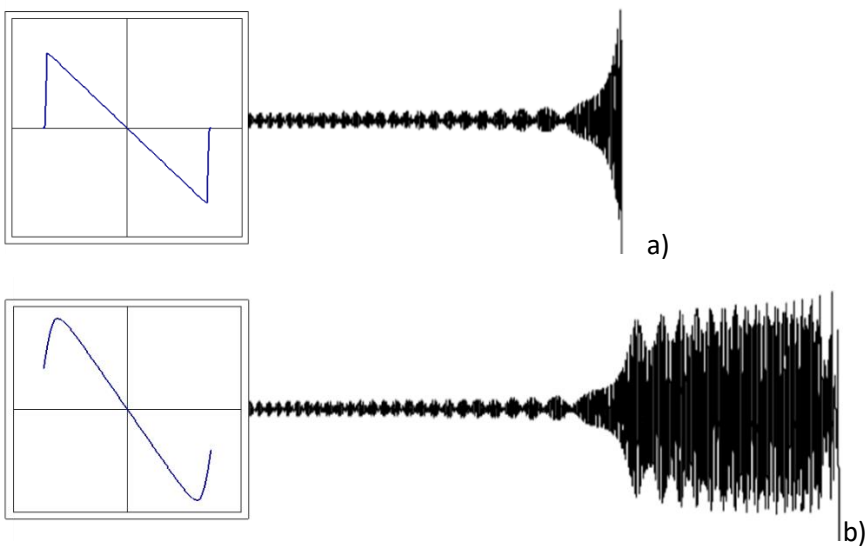

**Figure 4. Resonance ion ejection process: a) linear; b) nonlinear.**

Negative effects of the field distortion may compensate for each other to some extent, e.g. ejection holes decrease the field whereas electrode stretching increases the field. Mutual compensation results in more or less linear resonance, and rapid ejection. It should be emphasized here that the ideal quadrupole electric field is a sufficient but not necessary condition of the ion trap quality. The field inside Finnigan's 3D HIT (electrodes are holed and moved outward) is non-quadrupolar but resonance ion ejection is almost linear (low mass shift). The next step in solving this problem is to find a way to minimize the non-linearity. Some ion trap geometries (Finnigan's 3D HIT, Thermo's 2D HIT) have been experimentally proved to have smaller mass shifts than others (Rectilinear Ion Trap). Is it possible to find even better geometry minimizing non-linear effects arising during the ion ejection? Since the trial-and-error method is not the most efficient way for ion trap development, some theoretical approaches should be applied. There are three main approaches to the problem of resonant ion ejection: pure analytical solution, analysis of the multipole field expansion, and computer simulation.

## Analytical solution

Nonlinear resonance appears to be too complicated for a purely analytical solution. Only the simplest cases can be solved analytically. Let us consider how small field distortion affects resonance oscillation. Restoring force is supposed to deviate from the linear as a sinus of the displacement, $\frac{d^2x}{dt^2} = -\omega^2 \sin(x)$. Energy conservation law yields dependence of the eigenfrequency vs. oscillation amplitude $\omega_0(X)$ under the condition of small deviation, $\frac{\omega}{\omega_0} = \frac{2}{\pi} \int_0^1 \frac{dy}{\sqrt{(1-y^2)(1-\sin^2(X/2)y^2)}}$, where $y = \frac{\sin(x/2)}{\sin(X/2)}$. The solution is $\frac{\omega_0}{\omega} = \frac{\pi}{2K(X/2)}$, where $K$ is the complete elliptic integral of the first kind (tabulated function).

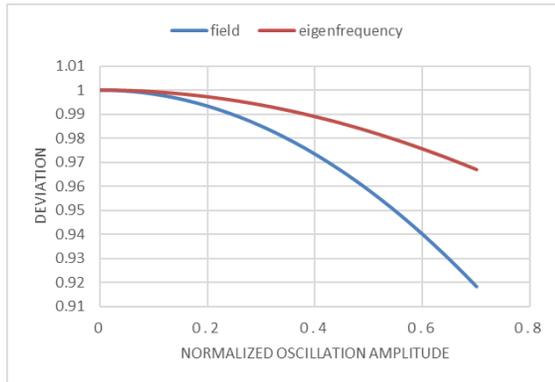

**Figure 5. Deviation of the electric field and eigenfrequency.**

The only conclusion that can be made is that the decreasing electric field (inevitable near the ejection holes) results in the decrease of the ion oscillation eigenfrequency. No practical suggestion can be made based on this information.

## Multipole potential expansion

Geometry other than hyperbolic, ejection holes, and the electrode's imperfection distort the field inside the IT. With field distortions, the potential is no longer exactly quadrupolar. Instead, the field is described as a superposition of multipoles[8], (quadrupole, hexapole, octopole, decapole, etc.): $\Phi(r,z,t) = (U + V\cos(\omega t)) \sum_{n=0}^{\infty} A_n f_n(r,z)$ , where $f_0(r,z) = 1$; $f_1(r,z) = \frac{z}{r_0}$; $f_2(r,z) = \frac{2z^2 - r^2}{2r_0^2}$; $f_3(r,z) = \frac{2z^3 - 3zr^2}{2r_0^3}$, etc. Non-quadrupolar field components result in the nonlinear field and therefore the nonlinear oscillations (oscillation eigenfrequency becomes amplitude-dependent). Because of the r-z cross-terms the ion motion in the x and z directions becomes mutually dependent. However, this effect is small since the cross-terms are of the second order of smallness. Some combinations of the multipole expansion coefficients appear to be better regarding the ion trap quality. The multipole expansion method can be used for the optimization of the ion trap geometry as follows:

- calculate the multipole field contributions associated with the initial ion trap geometry, and compare with a set of standard electric field charts to determine how the parameters need to be varied;
- adjust the parameters, calculate the electric fields, and repeat this process until an appropriate distribution of the electric field multipoles is obtained. The "-10%" compensation for positive octapolar and negative dodecapolar fields is suggested as a criterion for the appropriate field distribution for the resonance ejection mass scan;
- fabricate the ion trap with the optimized geometry and use it to record mass spectra for the experimental confirmation of the theoretical findings.

This empirical approach raises some questions. Is the account only the octapolar and dodecapolar sufficient? Does the "-10%" criterion provide the best possible MS quality? How many expansion coefficients are necessary for better predictability and reliability of the multipole expansion analysis?

### Computer modeling

Computer modeling is a virtual simulation of the ion movement inside the ion trap. That is a most straightforward approach including the following steps:

- Compute the electric field inside the trap;
- Simulate the ion movement near the stability boundaries;
- Adjust the ion trap geometry and repeat calculations until ion ejection becomes linear;
- Fabricate the ion trap with the optimized geometry and experimentally confirm the simulation.

The first step, the computation of the electric field distribution inside the trap is the easiest to accomplish. The solution of the Poisson equation for certain boundary conditions (determined by the electrode geometry) can be computed with high accuracy. But the second step, simulation of the ion ejection faces significant difficulties. Since the merit of the effect of interest (mass shift) is $10^{-3}...10^{-5}$, the accuracy of the ion movement simulation should be about the same orders of magnitude. Modern computer simulation software (SIMION, COMSOL, ITSIM) can't simulate ion movement with such high accuracy. The main problem here is the problem of relevance: how good is the correlation between real and simulated ion traps? So, experimental validation is a mandatory part of the solution of the resonance ion ejection problem.

The next section discusses the proposed semi-theoretical solution to the mass shift problem, validates it by comparison with known ion trap designs, and suggests alternative designs.

## Results and Discussion

This work aims to develop an ion trap mass analyzer that provides low mass shift at high scanning speed and is relatively easy to fabricate. This task may be divided into the following subtasks:

- Develop a theoretical procedure for the mass shift evaluation.
- Validate developed procedure on existing (known) ion traps.
- Find a simple but accurate ion trap design to solve the tolerance and trapping capacity problems.
- Apply the developed procedure to find ion trap geometry minimizing mass shift.

### Mass shift evaluation

Since mass shift results from the non-linear ion ejection, then to solve the problem one should find certain IT electrode geometry to minimize ejection resonance non-linearity. Experimental testing of all possible geometries is unrealistic, so the theoretical algorithm of the non-linearity evaluation is vital for solving the mass shift problem. The proposed algorithm of the non-linearity evaluation is based on the following notions:

- Ion ejection determines ion trap quality;
- Ion trap ejection is a resonance phenomenon;
- Resonant ion ejection should be linear for a low mass shift at high scanning speed;
- Low variation of the eigenfrequency in dependence on the ion oscillation amplitude is the sufficient and necessary condition of the linear resonance ejection.

The main idea of the proposed solution is that if the direct simulation of the resonance ejection is impossible, let us estimate resonance conditions and require them to be as linear as possible. So, the mass shift evaluation task may be reduced to minimizing the ejection resonance non-linearity for a given IT electrode geometry. Technically the evaluation procedure is implemented as follows:

- Compute field inside the trap of certain geometry.
- Compute free ion oscillation eigenfrequency in dependence of the oscillation amplitude in the pseudopotential well.
- Variation of the ion oscillation eigenfrequency in dependence of the ion oscillation amplitude is a criterion of the IT quality

The ion in the ion trap near the stability diagram border is affected by the following forces: the acceleration force is proportional to the ion mass; the "viscous" force (damping gas effect) is proportional to the ion velocity; the restoring force is proportional to the ion displacement (results from the movement in a pseudopotential well); the force due to the RF electric field is pure harmonic.

Summing all the forces affecting the ion yields the following ordinary differential equation $\ddot{x} + 2\zeta\omega_0\dot{x} + \omega_0^2 x = \frac{eE}{m}Cos(\omega t)$, where $m$ and $e$ are the ion mass and charge; $\varsigma$ is the damping coefficient proportional to the damping gas pressure; $E$ is RF field amplitude; $\omega_0$ is ion eigenfrequency. The steady-state solution ($t \to \infty$) can be written as $x(t) = A\cos(\omega t - \phi)$, where $\omega$ is the external force (RF field) frequency; $A = \frac{eE}{m\omega_0^2}\frac{1}{\sqrt{(1-\delta^2)^2 + (2\zeta\delta)^2}}$ is oscillation amplitude; $\delta = {}^{w}/_{w_0}$ is the external force frequency normalized on the eigenfrequency; and $\phi = arctan\left(\frac{2\zeta r}{1-\delta^2}\right)$ is a phase shift. However, since the eigenfrequency depends on the distance from the ion trap center only numerical methods should be used to solve the differential equation.

The transient solution of this equation describes the time evolution of the resonance oscillations amplitude, X(t). In the case of undamped resonance ($\varsigma = 0$ and $\omega = \omega_0$) the amplitude of harmonic oscillation increases to infinity linearly $X(t) = t\frac{eE}{2m\omega_0}$. In the case of damping resonance oscillations, the amplitude increases as $X(t) = A(1 - exp(-t\zeta\omega_0))$. So, the ejection time (i.e. time required to increase the oscillation amplitude from 0 to $x_0$) is $T(x_0) = -\frac{1}{\zeta\omega_0}ln\left(1 - \frac{x_0}{A}\right)$. Since the damping coefficient $\varsigma$ is small (A>>$x_0$) this equation can be linearized as $T(x_0) \approx \frac{1}{\zeta\omega_0}\frac{x_0}{A}$.

The proposed algorithm yields the normalized variation of the ion oscillation eigenfrequency in dependence on the normalized ion oscillation amplitude. Two additional steps are necessary for the correct quantitative estimation of the ion trap quality.

To compare different ion trap geometries, it was assumed that the ion approaching the ejection hole on the distance $d=d_0$ is captured by the dynode field and ejected. I.e. the range of the ion oscillation amplitude is equal to the distance from the ion trap center to the ejection hole minus half-width of the hole/slit, $x=x_0-d_0$. It corresponds to the range [0; 1] after normalization.

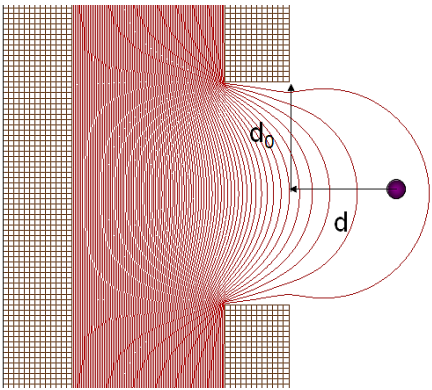

**Figure 6. Point of ejection**

For correct comparison and optimization mass shift should be characterized quantitatively. An ideal ion trap is a natural reference point for the ion trap characterization. The ratio of the ejection times of the ion trap under investigation ($T_1$) to the ideal ion trap ($T_0$) under the same conditions is proposed as a quantitative characteristic of the mass shift: $J = \frac{T_1}{T_0}$. The physical meaning of the mass shift coefficient $J$ is an ion ejection delay compared to the ideal ion trap (for the ideal IT $J=1$), so the smaller its value the higher the ion trap quality.

### Algorithm validation

The mass shift coefficient was calculated for the known ion trap designs for algorithm adjustment and validation. The following ion traps have been considered: 3D Hyperbolic Ion Trap (Finnigan), 2D Hyperbolic Ion Trap (Thermo), Rectilinear Ion Trap (Purdue), and Quadrupole Mass Filter.

*3D ion trap*

The electric field along the x-axis inside the ideal hyperbolic ion trap ($z_0/r_0$=0.708) without ejection holes is strictly linear; the pseudopotential well is parabolic; resonance ejection is linear. Ideal 3D HIT is a convenient object for estimating the algorithm's accuracy. To estimate overall computational error three calculations were performed starting from the electric field distribution along the x-axis obtained by:

1. Low-resolution SIMION electric field computation inside the ideal HIT (600 x 850 points array);
2. High-resolution SIMION electric field computation (3000x4000 field points array);
3. Strictly linear electric field (defined analytically).

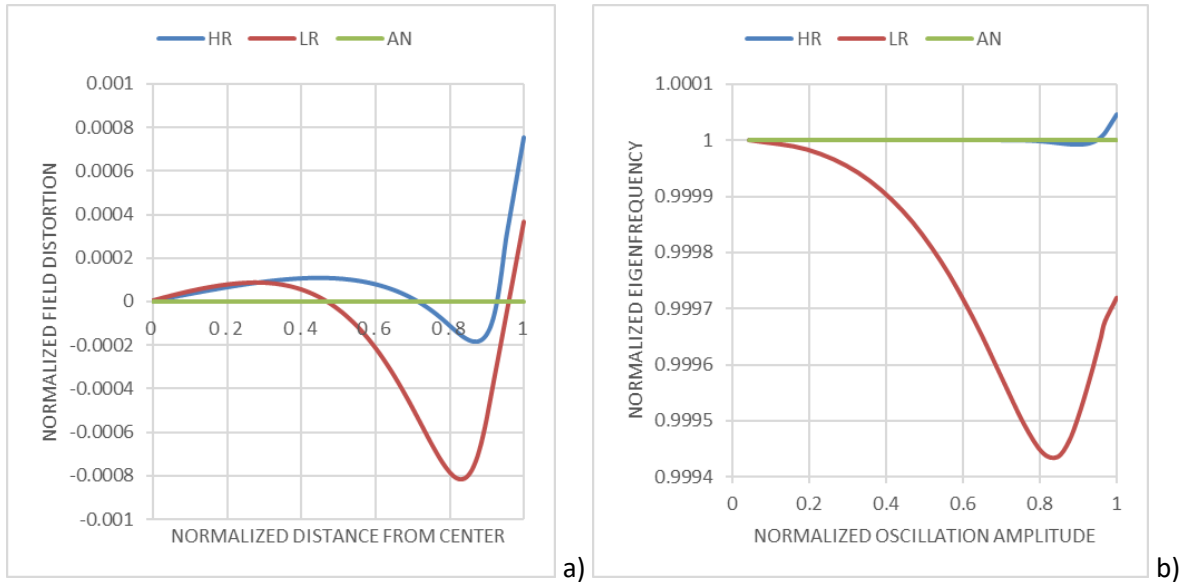

a)  b)

**Figure 7. Ideal 3D Hyperbolic Ion Trap computations: a) field distortion; b) eigenfrequency variation. (HR - High-Resolution field; LR - Low-Resolution field; AN – analytically defined field)**

Corresponding mass shift coefficients are $J_{LR}$=1.49, $J_{HR}$=1.024, and $J_{AN}$=1. The overall accuracy of the algorithm is high given the typical range of the mass shift coefficients [1…20]. Also, one can conclude that the main error source is field computation.

To have any practical value IT must have the ejection holes. Ejection holes disturb electric field distribution inside the IT, so the field is non-linear, the pseudopotential well is not parabolic, and the resonance ejection is non-linear. Electric field strength near the ejection holes in the endcaps is smaller than the linear electric field in the ideal trap (without holes). The electric field strength near the endcap moved outward is larger than the linear electric field in the ideal trap (without moving). Combining both ejection holes and endcaps moving is expected to result in the mutual compensation of the negative effects to some extent. It had been found experimentally that moving endcaps apart without reshaping (the electrode shape is given by condition $z_0/r_0$=0.707 but the distance between the ion trap center and the endcaps is *0.783$r_0$*) results in the smallest mass shift[9]. We apply our algorithm to all these cases to compare our theory with the experiment. The comparison includes geometry optimization, i.e. search for the best possible HIT geometry minimizing mass shift. Hyperbolic Ion Trap geometry is determined by just one parameter, the half-distance between holed electrodes normalized on the ring electrode radius $z_0/r_0$. So, the geometry optimization is pretty straightforward – the minimum of the optimization function $J(z_0/r_0)$ corresponds to the best HIT. The following 3D Hyperbolic Ion Traps are examined:

1. Ideal HIT (IH) with holes: mass shift coefficient was calculated as a starting point;
2. Moved HIT with holes (MH): endcaps are moved outward without reshaping; geometry parameter range is $z_0/r_0$ = [0.7 – 0.8];
3. Stretched HIT with holes (SH): the endcaps are reshaped so the electrode profile satisfies the hyperbolic equation; the geometry parameter range is $z_0/r_0$ = [0.7 – 0.8].

For the stretched HIT, the optimal geometry parameter is $z_0/r_0 = 0.74$, and the mass shift coefficient is $J_{SH}(0.74)=5.87$. For the moved HIT, the optimal geometry parameter is $z_0/r_0 = 0.78$, and the mass shift coefficient is $J_{MH}(0.78)=4.02$. The calculated value of *0.78* agrees with the experimental value of 0.783.

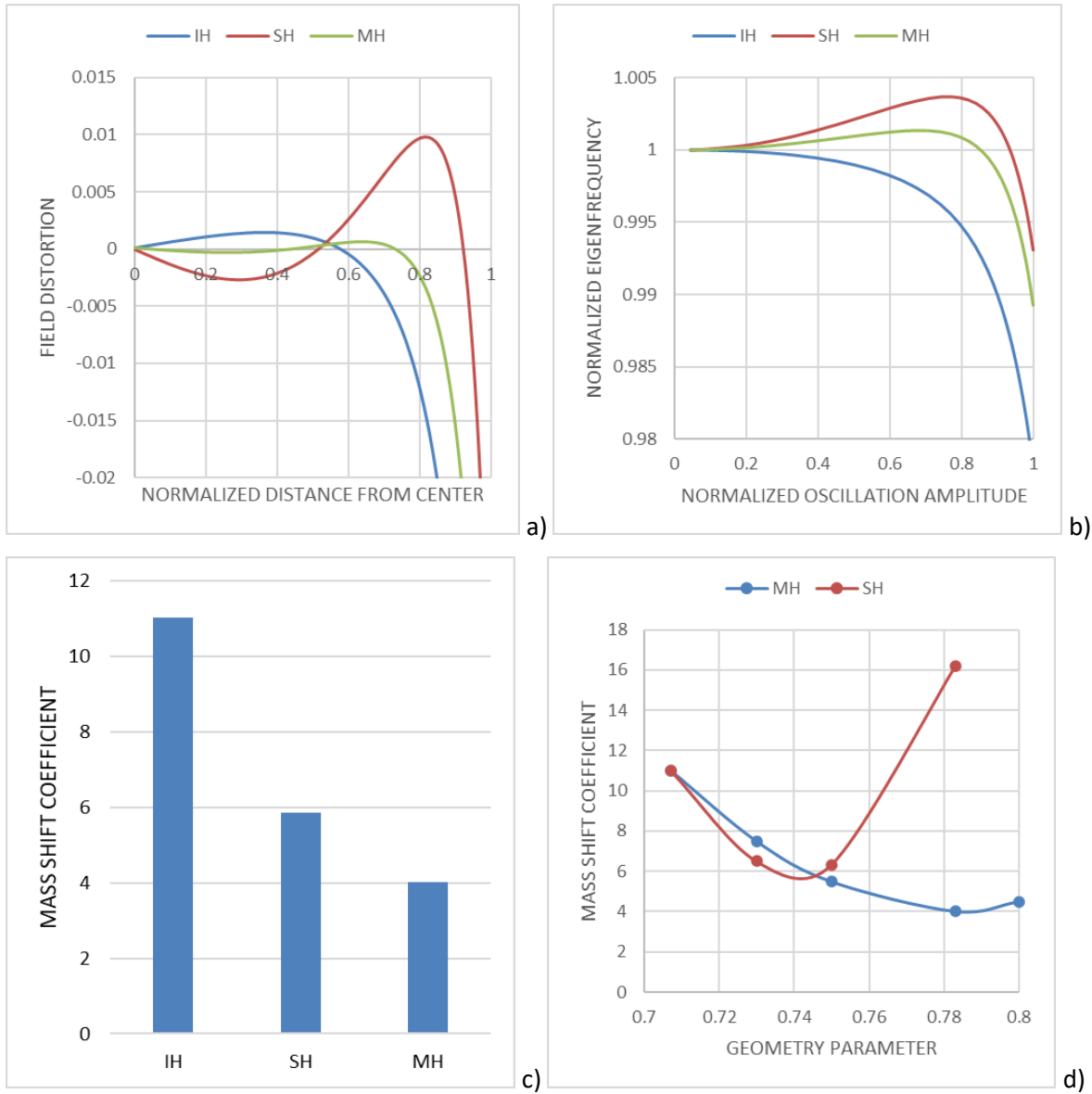

**Figure 8.** a) Field distortion; b) Eigenfrequency vs. oscillation amplitude; c) Mass shift coefficient for the 3D ion trap; d) HIT optimization. (IH – ideal with holes; SH – stretched with holes; MH – moved with holes).

### *2D ion trap*

Linear (2D) hyperbolic Ion Trap (LIT) (proposed for larger trapping capacity and higher trapping efficiency) faces the same mass shift problem as a 3D Hyperbolic Ion Trap (HIT). This problem for LIT had been solved in the same way as for HIT: moving apart slated electrodes to compensate for electric field distortion due to the ejection slits[10]. We optimize LIT geometries and compare our results with known experimental data. Linear Ion Trap geometry is determined by one parameter - the distance between slotted x-electrodes normalized on the distance between y-electrodes, $x_0/y_0$. The minimum of the optimization function $J(x_0/y_0)$ corresponds to the best LIT. We apply our algorithm to the following 2D Ion Traps:

1. Ideal LIT with slits (IH): mass shift coefficient was calculated as a starting point;
2. Moved LIT with slits (MH): slotted endcaps are moved outward without reshaping; geometry parameter range is $x_0/y_0 = [1 - 1.2]$;

3. Stretched LIT with slits (SH): slotted endcaps are reshaped so the electrode profile satisfies the hyperbolic equation; geometry parameter range is $x_0/y_0 = [1 - 1.2]$;
4. Rectilinear Ion Trap (RH); geometry parameter range is $x_0/y_0 = [1 - 1.45]$;

For the stretched LIT optimal $x_0/y_0 = 1.05$, and the mass shift coefficient is $J_{SH}(1.05)=9.89$. For the moved LIT optimal $x_0/y_0 = 1.15$, and the mass shift coefficient is $J_{MH}(1.15)=4.86$. A mass shift for the Rectilinear Ion Trap (developed to circumvent manufacturing difficulties that hinder LIT) appears unavoidable. Even in the optimal point ($x_0/y_0=1.25$) the mass shift coefficient $J_{RIT}(1.25)=14.9$ is much larger than for LIT. That is consistent with practice: the analytical quality of RIT is lower than LIT. The calculated values of the optimal geometry parameters for both LIT and RIT coincide with the experimentally found values.

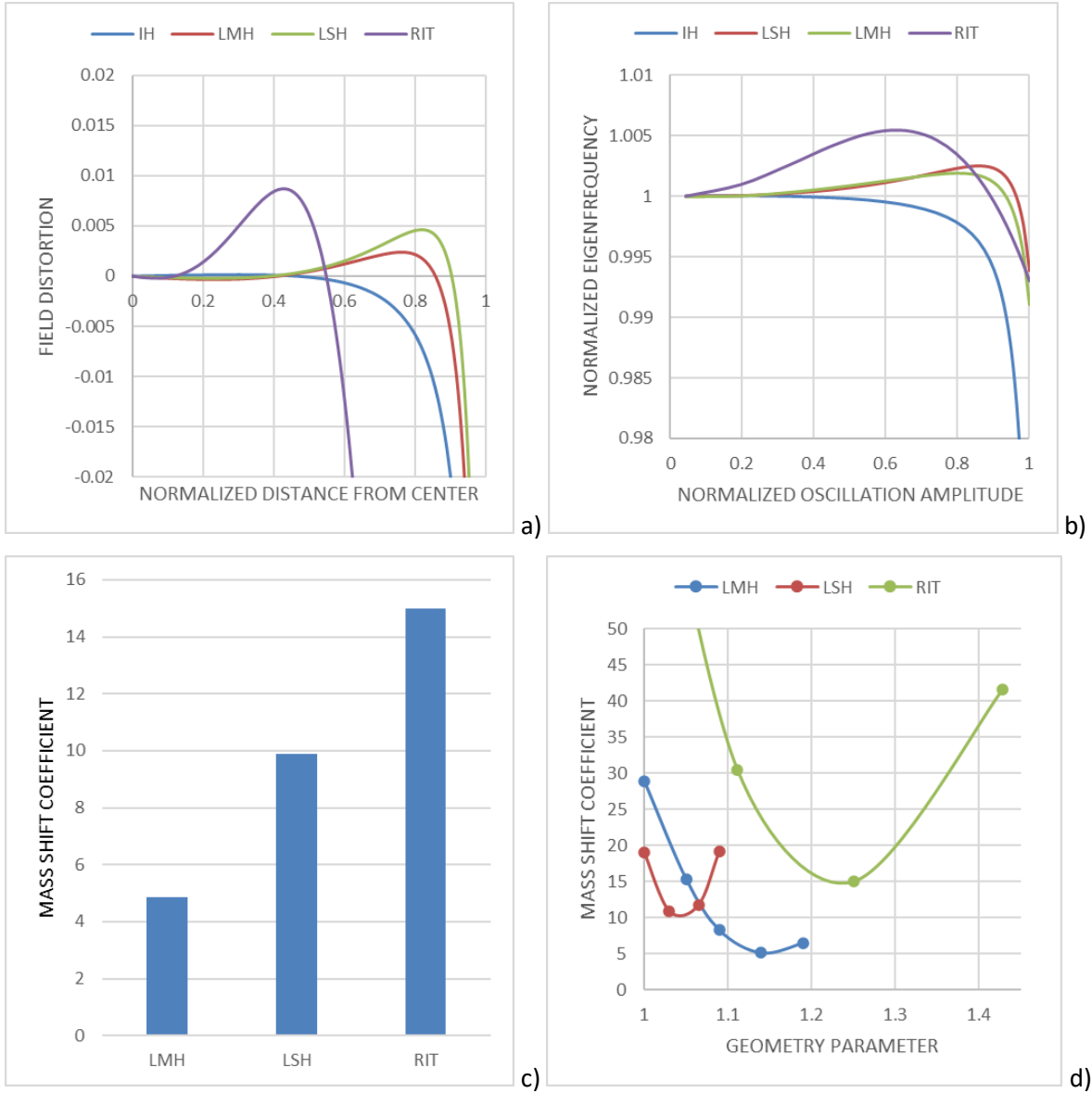

**Figure 9.** a) Field distortion; b) Eigenfrequency vs. oscillation amplitude; c) Mass shift coefficient for the 2D ion trap; d) IT optimization. (IH – ideal with holes; LSH – stretched with holes; LMH – moved with holes; RIT - rectilinear).

Thus, we can conclude that the experiment validates the proposed algorithm of the mass shift evaluation because the predicted optimal geometry of the ion traps is the same as found experimentally. The next step is to find the ion trap geometry of low mass shift, simple in fabrication, and satisfying the requirement of high mechanical accuracy.

## Alternative geometries

The search for the right design for a high-quality Ion Trap was based on the following facts:

- 2D IT has the advantage of higher ion capacity over 3D IT.
- Round or flat rods are the easiest for accurate fabrication.

So, for the best quality, IT should be linear with flat or round electrodes. Two classes of electrode geometry (round and triangular) appear to be the most promising for high-quality IT. The mass shift evaluation algorithm was applied to the chosen class of IT geometries to find optimal parameters minimizing mass shifts.

### *Round electrodes*

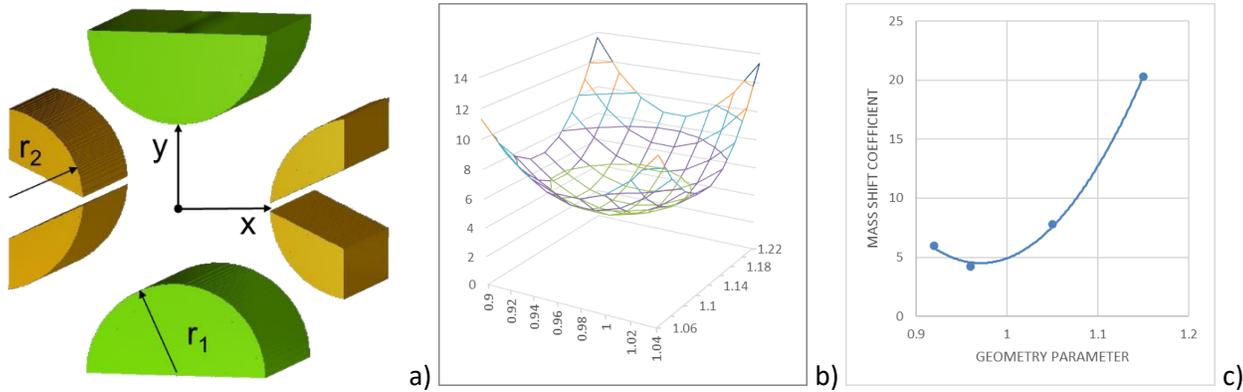

**Figure 10. Round Rods Linear Ion Trap: a) geometry, axes, and optimization parameters; b) local minimum in the parametrical space; c) optimization curve for QMF.**

Linear ion trap with round rods is a challenging optimization task. There are three optimization parameters for the cylindrical rods geometry normalized on the half-distance between non-slotted rods ($y_0$): slotted rods radius ($r_2/y_0$), non-slotted rods radius ($r_1/y_0$), and the half-distance between slotted rods ($x_0/y_0$). Normalized ejecting slits ($d/y_0$) were 0.075, which corresponds to 0.3 mm for $y_0=4$ *mm*. Search for a global optimum in the multidimensional parametrical space is an extremely time-consuming problem. However, finding the local optimum is possible, resulting in the decent mass shift factor *J(0.97)=4.5* (compared with *J=4.86* for LIT). There is no need to mention that cylindrical electrodes are much easier to fabricate with the required accuracy than hyperbolic electrodes.

Without ejection slits, Round Rods Linear Ion Trap becomes a Quadrupole Mass Filter. QMF geometry is much easier for optimization. Symmetry considerations ($r_1=r_2=r$, $x_0=y_0=r_0$) reduce the number of the optimization parameters to one: the ratio of the rod's radius and the half-distance between them $r/r_0$. The mass shift evaluation algorithm predicts an optimum for the Quadrupole Mass Filter, $J_{QMF}$*(1.13)=1.94*, which agrees with standard quadrupole MS geometry ($r/r_0$= *1.1487*).

### *Triangular rods*

Optimization parameters for the triangular IT are the half-distance between slotted rods ($x_0$), the half-distance between non-slotted rods ($y_0$), the width (*A*) and height of the non-slotted rods (*a*), and the width (*B*) and height of the slotted rods (*b*). The scaling principle allows for reducing the number of the variables by one, from six to five. For certain values of the optimization parameters the mass shift coefficient appears to be extremely small, *J(1.03)=2.36*.

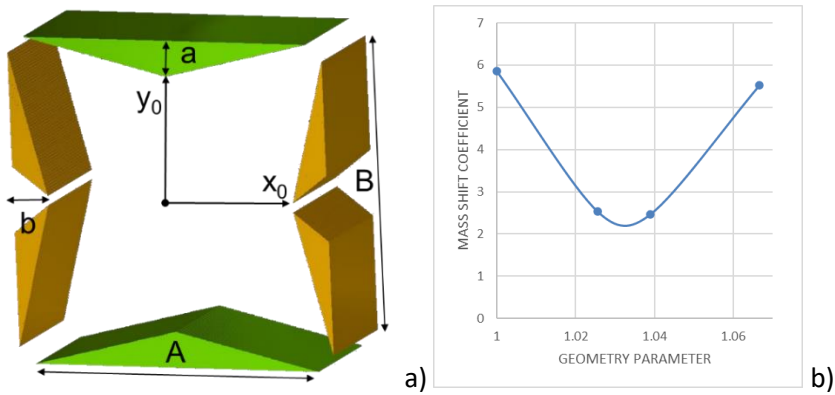

**Figure 11. Triangular Rods Linear Ion Trap: a) geometry, axes, and optimization parameters; b) local minimum in the parametrical space.**

## Conclusion

This work aims to develop a high-quality ion trap. The following subtasks are solved: the theoretical procedure of the mass shift evaluation is developed; this procedure is extensively validated by the existing (known) ion traps; simple in fabrication but accurate ion trap geometries are proposed; these geometries are optimized in terms of low mass shift (and therefore highest IT quality) by the developed theoretical procedure.

The developed theoretical procedure of the mass shift evaluation:

- proposes a criterion of the ion trap evaluation: eigenfrequency in dependence of the ion oscillation amplitude;
- computes this dependence with high accuracy (<0.05%);
- allows correct quantitative IT quality comparison for different IT geometries;
- has passed validation test: simulation results are consistent with existing trap designs (3D hyperbolic IT, 2D hyperbolic IT, Rectilinear IT, and Quadrupole Mass Filter);
- can be used for theoretical optimization of the ion trap geometry.

As a result of the application of this procedure to various IT geometries, it has been found that in terms of the mass shift:

- Linear ion trap with triangle rods appears to be the best (mass shift factor J=2.36, flat electrodes can be fabricated with high accuracy).
- Linear ion traps with round rods (J=4.5) are better than linear ion traps with hyperbolic rods (J=4.86) and are easier to fabricate.
- Existing ion traps (QMF, HIT, LIT, RIT) are optimal in a framework of chosen geometry.

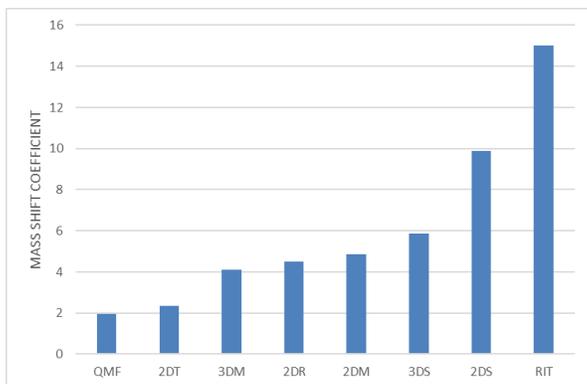

**Figure 12. Mass shift coefficient for Quadrupole Mass Filter (QMF), 2D IT with triangle rods (2DT), 3D IT with moved hyperbolic electrodes (3DM), 2D IT with round rods (2DR), 2D IT with moved hyperbolic electrodes (2DM), 3D IT with stretched hyperbolic electrodes (3DS), 2D IT with stretched hyperbolic electrodes (2DS), and Rectilinear IT (RIT).**

# References

[1] Paul, Wolfgang; Steinwedel, Helmut (1953). *Zeitschrift für Naturforschung A*. **8** (7): 448–450. doi:10.1515/zna-1953-0710

[2] R.E. March, J.F.J. Todd; *Quadrupole ion trap mass spectrometry*, 2nd ed.; Wiley, 2005; p 83.

[3] US Patent 5,420,425; Bier M E., Syka J E.; *Ion trap mass spectrometer system and method*, May 30, 1995, Filed: May 27, 1994

[4] Titov VV. Int. J. *Mass Spectrom. Ion. Processes* 1995; 141: 37.

[5] Dawson PH. Adv. *Electron. Electron Optics* 1980; 53: 153

[6] Ouyang, Z. W., G.; Song, Y.; Li, H.; Plass, W. R.; Cooks, R. G.; *Anal. Chem.* 2004, *76*,4595-4605.

[7] Wolfgang R. Plass, Hongyan Li, R. Graham Cooks, *International Journal of Mass Spectrometry*, 228 (2003) 237–267

[8] Guangxiang Wu, R. Graham Cooks, Zheng Ouyang, *International Journal of Mass Spectrometry,* 241 (2005) 119–132

[9] Syka, J. E. P. Commercialization of the Quadrupole Ion Trap. March, R. E.; Todd, J. F. J., Eds. Practical Aspects of Ion Trap Mass Spectrometry, Volume 1: Fundamentals of Ion Trap Mass Spectrometry, 1. CRC Press: Boca Raton, FL, 1995; 169–205.

[10] J C. Schwartz, M W. Senko, J E. P. Syka, *J Am Soc Mass Spectrom*, 2002, 13, 659–669